\documentclass{aastex}          
\usepackage{spr-astr-addons}    

\begin{document}
%
\title{On the horizons in a viable vector-tensor theory of gravitation}

\shorttitle{<Horizons in a vector-tensor theory>}
\shortauthors{<R. Dale, M. Fullana and D. S\'aez>}

\author{Roberto Dale\altaffilmark{1}}
\affil{Departament d'Estad\'{\i}sica, Matem\`atiques i Inform\`atica, 
Universitat Miguel Hern\'andez, 03202-Elx, Alacant, Spain.}
\email{rdale@umh.es} 
\and
\author{M\`{a}rius J. Fullana\altaffilmark{2}}
\affil{Institut de Matem\`atica Multicisciplin\`aria, Universitat
Polit\`ecnica de Val\`encia, 46022, Val\`encia, Spain}
\email{mfullana@mat.upv.es} 
\and  
\author{Diego S\'aez\altaffilmark{3}}
\affil{Departamento de Astronom\'\i a y Astrof\'\i sica, 
Universidad de Valencia, 46100-Burjassot, Valencia, 
Spain, and Observatorio Astron\'omico, Universidad de Valencia, E-46980 Paterna, Valencia, Spain}
\email{diego.saez@uv.es} 

\begin{abstract} 
A certain vector-tensor (VT) theory of gravitation was tested in previous papers. 
In the background universe, the vector field of the theory has a certain energy density,
which is appropriate to play the role of vacuum energy (cosmological constant). Moreover, 
this background and its perturbations may explain
the temperature angular power spectrum of the cosmic microwave background (CMB)
obtained with WMAP (Wilkinson Map Anisotropy Probe), and other observations, as e.g., the Ia supernova 
luminosities. The parametrized post-Newtonian limit of the VT theory has been proved to be 
identical to that of general relativity (GR), and there are no quantum ghosts
and classical instabilities.
Here, the stationary spherically symmetric solution, in the absence of any 
matter content, is derived and studied. The metric of this solution 
is formally identical to that of the Reissner-Nordstr\"{o}m-de 
Sitter solution of GR, but the role of the electrical charge 
is played by a certain quantity $\Gamma $ depending on both the vector field and the 
parameters of the VT theory. The black hole and cosmological horizons are 
discussed. The radius of the VT black hole horizon deviates with respect to 
that of the Kottler-Schwarzschild-de Sitter radius. Realistic relative deviations
depend on $\Gamma $ and reach maximum values close to 30 per cent. 
For large enough $\Gamma $ values, there is no any black hole horizon, 
but only a cosmological horizon. 
The radius of this last horizon is almost independent of the mass source, the vector field 
components, and the 
VT parameters. It essentially depends on the cosmological constant value,
which has been fixed by using cosmological observational data (CMB anisotropy, 
galaxy correlations and so on).  

\end{abstract}


\section{Introduction.}
\label{sec-1}
Recently, several vector-tensor (VT) theories --involving a vector field, $A^{\mu}$,
and the metric tensor $g^{\mu \nu}$-- have been applied to cosmology \citep{dal09,dal12}; 
in these theories, the background energy density, $\rho^{A}_{B}$, of the vector field $A^{\mu}$ plays the role of the dark energy
(hereafter the subscript $B$ stands for background);
for example, in \citet{dal09}, where the theory of gravitation considered in this paper was proposed,
the equation of state is $P^{A}_{B} = W \rho^{A}_{B} $, where $P^{A}_{B}$ is the pressure due to the 
field $A^{\mu}$ and $W =-1 $; hence, the constant energy $\rho^{A}_{B} $ plays the role of 
vacuum energy. However, in the theory studied in \citep{dal12}, which might be appropriate to 
explain the anomalies observed in the angular spectrum of the cosmic microwave background (CMB) 
for small $\ell$ multipoles, 
the equation of state is $P^{A}_{B} = W(a) \rho^{A}_{B} $, where $W(a)$ is negative for any value of the scale factor $a$; 
hence, in this theory, we have a sort of dynamical dark energy different
from that associated to the cosmological constant (vacuum energy).

Here, our attention is focused on the theory proposed by \citet{dal09}, which was applied to cosmology 
in \citet{dal12prd} and \citet{dal14}. In this last reference, the VT theory under consideration
was proved to be viable in the sense that: (i)
its post-Newtonian parametrized limit is identical to that of general relativity (GR), and (ii) 
the theory may simultaneously explain the seven year WMAP data about the 
CMB temperature anisotropy and the measurements of supernova Ia luminosities.
Conclusion (ii) was obtained by using 
the well-known Bardeen formalism \citep{bar80} to
write the evolution equations of the scalar linear perturbations in VT theory, and also to find
the initial conditions 
at high redshift necessary to solve these equations [see \citet{mab95}]. By using these elements, a 
modified version of the code COSMOMC \citep{leb02} --based on statistical techniques as the Markov chains -- 
was designed to fit the VT predictions  
with the observational data mentioned above. 
A model involving seven free cosmological parameters was used.
Results were encouraging \citep{dal14} and the theory deserves attention.

Before writing any field or cosmological equation, let us fix some notation criteria. 
Our signature is (--,+,+,+). Latin (Greek) 
indexes run from 1 to 3 (0 to 3).The symbol $\nabla $ ($\partial $) stands
for a covariant (partial) derivative. The antisymmetric tensor $F_{\mu \nu} $
is defined by the relation 
$F_{\mu \nu} = \partial_{\mu} A_{\nu } - \partial_{\nu} A_{\mu }$,
in terms of the $A^{\mu }$ vector field.
Quantities $R_{\mu \nu}$, $R$, and $g$ are the covariant components of the Ricci 
tensor, the scalar curvature and
the determinant  of the matrix $g_{\mu \nu}$ formed by the covariant components 
of the metric, respectively. The gravitational constant
is denoted $G$ and the speed of light $c$. 
Units are chosen in such a way that $c= G = 1$; namely, we use 
geometrized units. The dimension of any quantity is $L^{n} $, n being an integer number.
Length unit is chosen to be the kilometer.
Our coordinates are denoted $t$, $r$, $\theta $, and $\phi$. Whatever the quantity 
$\xi $ may be, $\xi^{\prime} $ stands for a partial derivative with respect to the 
radial coordinate $r$. 

The two VT theories mentioned above correspond to different choices of the parameters
$\omega$, $\eta $, $\varepsilon$,
and $\gamma$ 
involved in the action \citep{wil93}:
\begin{eqnarray}
I &=& \int (  R/16\pi + \omega A_\mu  A^\mu  R
+ \eta R_{\mu \nu }
A^\mu  A^\nu  -  \nonumber \\
& &
\varepsilon F_{\mu \nu } F^{\mu \nu }
+\gamma \,\nabla_\nu  A_\mu  \nabla^\nu  A^\mu
+ L_{m} ) \,\sqrt { - g} \,d^4 x \ ,
\label{1.1}
\end{eqnarray}  
where the tensor $F_{\mu \nu} $ --defined above--
is not the electromagnetic one. The VT theory studied in this paper [see \cite{dal09}] corresponds to the following 
choice of the free dimensionless parameters involved in action (\ref{1.1}):
$\omega=0 $ and $\eta = \gamma $. In this theory of gravitation, it has been proved 
that there are no ghosts and unstable modes 
for $2\varepsilon - \gamma > 0$. 
Moreover, for a homogeneous and isotropic Robertson-Walker background universe, the energy density 
of the vector field $\rho^{A}_{B}$ has been proved to be 
\begin{equation}
\rho^{A}_{B} = \gamma (\nabla \cdot A)_{B}^{2} \ ,
\label{eqest_vt}
\end{equation} 
where $(\nabla \cdot A)_{B} = (\nabla_{\mu} A^{\mu})_{B} $. 
Therefore, constant $\gamma $ must be positive to have $\rho^{A}_{B} >0$. 
The vector field equations, when applied to  Robertson-Walker cosmology, predict a constant value for 
$(\nabla \cdot A)_{B}$ and, consequently, $\rho^{A}_{B}$ is strictly constant
as the vacuum energy density. The viability of VT as 
a theory of gravitation and its cosmological success 
require $\gamma $  and $\varepsilon $ parameters satisfying the
inequalities $\varepsilon > \frac {\gamma}{2}>0 $, but their values cannot be fixed.
In this situation, it is worthwhile the design of new 
applications of the VT theory with the essential aim of fixing $\gamma $ and $\varepsilon $ 
and any other arbitrary quantity related with our limited knowledge 
of the $A^{\mu}$ nature and properties, which are being analyzed. Previous outcomes (based on linearity) strongly suggest that 
the new applications should be nonlinear.    
On account of these considerations, we have planed the study of various 
gravitational physical systems as, for example:
(a) the black hole horizons of different sizes and their neighborhoods, (b) 
the cosmological evolution of nonlinear 
structures (galaxies, clusters, superclusters and so on) by using
either approximations or simulations, and (c) binary stellar systems radiating 
gravitational waves.
Here, our attention is focused on the simplest of these problems:
the study of the VT horizons of outer (no matter content) spherically symmetric 
stationary space-times, which are fully characterized by the mass $m$
(no electrical charge and rotation) and the cosmological constant $\Lambda $.

According to its formulation, VT is a theory of pure gravitation. 
The field $A^{\mu} $ has nothing to do 
with the potential vector of the electromagnetic field. It does not couple with 
electrical currents. The U(1) gauge 
symmetry of Maxwell theory is not required in VT. In other words, 
VT is a simple and manageable theory of gravitation. 
The electromagnetic interaction must be described in the standard way.

There are many alternative theories which are being currently studied,
some of these theories are only concerned with gravitation; e.g., the so-called $f(R) $ and 
$f(T)$ theories, where $T$ is
the torsion scalar. In these theories the electromagnetic field 
is treated in the standard way (minimal coupling with the gravitational part of
the Lagrangian).
In other theories the action is designed to describe both 
the gravitational and the electromagnetic fields; interesting cases may 
be found, e.g., in \citet{nov08}, where non-minimal couplings of the electromagnetic field with 
gravity are proposed. A very promising non-minimal coupling between 
the electromagnetic field and a $f(R)$ function is applied to cosmology in \citet{bam08},
where it is claimed that the theory is viable, and also that 
inflation and late time acceleration may be simultaneously 
explained. However, in VT theory, as well as in GR, it must be recognized that 
inflation is to be produced by additional fields. It is due to the fact that 
inflation must lead to an isotropic universe, whereas the inflation due to 
a vector field is expected to be anisotropic. 
Only a triplet of orthogonal vector fields or N randomly oriented vector fields
might produce an isotropic enough expansion \citep{gol08}, but this is 
not the case of the VT theory.

In order to explain inflation we could replace $R$ by an
appropriated function $f(R) $ in the Lagrangian of the VT theory; in this way,
the field $A^{\mu} $ could explain the accelerated late time expansion, whereas 
the scalar field, associated to $f(R)$ in the Einstein frame, could account for the 
required inflation; hence, function $f(R) $ would be chosen to achieve a good 
inflation, without producing late time acceleration, which implies less restrictions to 
be satisfied by f(R). Nevertheless, we think that before any generalization,
the VT theory must be fully developed as a simple viable and manageable 
gravitation theory, which explains many observations (see above) for arbitrary values 
of $\epsilon $ and $\gamma $. 

This paper is structured as follows: The VT theory is described in Sect. \ref{sec-2}, 
the stationary spherically symmetric solutions of the field equations are found in Sect. \ref{sec-3},
the horizons are studied in Sect. \ref{sec-4} and; finally, 
Section \ref{sec-5} 
contains a general discussion 
and a summary of the main conclusions.

\section{The VT theory: basic equations.}   
\label{sec-2}

Variational calculations based on action (\ref{1.1}), 
with $\omega=0 $ and $\eta = \gamma $, 
lead to the following field
equations \citep{dal14}:   

\begin{equation}  
G^{\mu \nu} = 8\pi  \, (T^{\mu \nu}_{m} + T^{\mu \nu}_{VT}) + T^{\mu \nu}_{\Lambda} \ ,
\label{fieles_vt}
\end{equation} 
where $T^{\mu \nu}_{m}$ is the contribution of matter to the  
energy-momentum tensor,
which have the same form as in GR.
Tensor 
$T^{\mu \nu}_{VT}$ is the contribution due to the 
$A^{\mu} $ field of the VT theory, whose form is
\begin{eqnarray}
T^{\mu \nu}_{VT} &=& 2(2\varepsilon - \gamma) [F^{\mu}_{\,\,\,\, \alpha}F^{\nu \alpha}
- \frac {1}{4} g^{\mu \nu} F_{\alpha \beta} F^{\alpha \beta}] \nonumber \\ 
& &
-2\gamma [ \{A^{\alpha}\nabla_{\alpha} (\nabla \cdot A) + \frac {1}{2}(\nabla \cdot A)^{2}\}
g^{\mu \nu}\nonumber \\ 
& &
-A^{\mu}\nabla^{\nu} (\nabla \cdot A) - A^{\nu}\nabla^{\mu} (\nabla \cdot A)
] \ ,
\label{emtee_vt}
\end{eqnarray}    
and $T^{\mu \nu}_{\Lambda} $ also have the same form as in GR; namely, 
\begin{equation}
T^{\mu \nu}_{\Lambda} = - \Lambda g^{\mu \nu} \ .
\label{tmnl}
\end{equation}

Eqs.~(\ref{fieles_vt}) are a generalization of the Einstein equations of GR.

Variations of the vector field $A^{\mu}$ in action (\ref{1.1}) give
\begin{equation}
2(2\varepsilon - \gamma)\nabla^{\nu} F_{\mu \nu} = J^{^{A}}_{\mu} \ ,
\label{1.3_vt}
\end{equation}     
where $J^{^{A}}_{\mu} = -2 \gamma \nabla_{\mu} (\nabla \cdot A)$ plays the role of a fictitious current. 

From Eq.~(\ref{1.3_vt}) one easily gets the relation 
\begin{equation}
\nabla^{\mu} J^{^{A}}_{\mu} = 0 \ , 
\label{confic}
\end{equation} 
which may be seen as the conservation law of the fictitious current $J^{^{A}}_{\mu}$
defined above.

Since the parameters $\gamma $ and $\epsilon $ are dimensionless,
a dimensional analysis of Eqs.~(\ref{fieles_vt}) and~(\ref{emtee_vt})
leads to the important conclusion that the
dimension of the $A^{\mu} $ components is $L^{0} $. This fact will 
be important below. By using the chosen units and the relation between the 
Einstein ($G^{\mu \nu} $) and Ricci ($R^{\mu \nu} $) tensors, Eqs.~(\ref{fieles_vt})
may be written as follows:
\begin{equation} 
R^{\mu}_{\,\,\,\nu} - \frac{1}{2} R \delta^{\mu}_{\,\,\,\nu} \ = \ T^{\mu}_{\,\,\,\nu} \ ,  
\label{gt} 
\end{equation} 
where $\delta^{\mu}_{\,\,\,\nu}$ is the Kronecker delta, and 
\begin{equation} 
T^{\mu \nu} =  8\pi  \, (T^{\mu \nu}_{m} + T^{\mu \nu}_{VT}) + T^{\mu \nu}_{\Lambda} \ .
\label{tmn} 
\end{equation} 
Equation~(\ref{gt}) may be easily 
rewritten in the form:
\begin{equation}
T^{\mu}_{\,\,\,\nu} - \frac{1}{2} T \delta^{\mu}_{\,\,\,\nu} \ = \ R^{\mu}_{\,\,\,\nu} \ ,
\label{tg}
\end{equation} 
$T$ being the scalar $T^{\mu \nu}g_{\mu \nu}$.

We have the basic equations to look for horizons in next Sections.

\section{The stationary spherically symmetric case in the VT theory}   
\label{sec-3}

It is well known that, in the stationary spherically symmetric case, the line element
may be written as follows [see e.g., \cite{ste03}]:
\begin{eqnarray}
ds^{2}&=&\ - e^{2\alpha(r)} d\tau^2 + e^{2\beta(r)} dr^2 \nonumber \\
& &
+ r^2 ( \ d\theta^2 + sin^2\theta \ d\phi^2) \ , 
\label{LE}
\end{eqnarray}
and, moreover, the covariant components $A_{\mu}$ have the form:
\begin{equation}
A_{\mu} \ \equiv \ [A_0(r),A_1(r),0,0] \ .
\label{VF}
\end{equation}
Accordingly, the nonvanishing $F_{\alpha \beta}$ components are
\begin{equation}
F_{10} = -F_{01} = A^{\prime}_{0}  \ .
\label{nvf}
\end{equation}

We have the four unknown functions $\alpha(r)$, $\beta(r)$, $A_0(r)$, and $A_1(r)$ to be found 
from the field equations of Sect.~\ref{sec-2}.

Hereafter, it is assumed that the matter tensor $T^{\mu \nu}_{m} $ vanishes and, then, 
taking into account Eqs.~(\ref{emtee_vt}), (\ref{tmnl}), (\ref{tmn}), 
plus Eqs.~(\ref{LE})--(\ref{nvf}), one easily get that, 
in terms of the new dimensionless parameters 
$\tilde{\gamma} = 8\pi \gamma $ and $\tilde{\varepsilon} = 8\pi \varepsilon $,
the nonvanishing $T^{\mu \nu}$ components are:
\begin{eqnarray}
T^{01} &=& T^{10} =  2 \tilde{\gamma} g^{00} g^{11} A_0 (\nabla \cdot A)^{\prime} \ ,
\label{TA01}
\end{eqnarray} 
\begin{eqnarray}
T^{00}&=&\  g^{00} \{ (2 \tilde{\varepsilon} - \tilde{\gamma})  [ g^{00} g^{11} (A^{\prime}_{0})^2]\nonumber \\ 
& &
-\tilde{\gamma}[2 A^1  (\nabla \cdot A)^{\prime} + (\nabla \cdot A)^2 ] - \Lambda \} \ ,
\label{T00}
\end{eqnarray}
\begin{eqnarray}
T^{11}&=&\  g^{11} \{ (2 \tilde{\varepsilon} - \tilde{\gamma})  [ g^{00} g^{11} (A^{\prime}_{0})^2]\nonumber \\ 
& &
-\tilde{\gamma}[-2 A^1  (\nabla \cdot A)^{\prime} + (\nabla \cdot A)^2] - \Lambda \} \ ,
\label{T11}
\end{eqnarray}
\begin{eqnarray}
T^{22}&=&\  g^{22} \{ (\tilde{\gamma} - 2 \tilde{\varepsilon} )  [ g^{00} g^{11} (A^{\prime}_{0})^2]\nonumber \\ 
& &
-\tilde{\gamma}[2 A^1  (\nabla \cdot A)^{\prime} + (\nabla \cdot A)^2] - \Lambda \} \ ,
\label{T22}
\end{eqnarray}
\begin{eqnarray}
T^{33}&=& \  g^{33} \{ (\tilde{\gamma} - 2 \tilde{\varepsilon} )  [ g^{00} g^{11} (A^{\prime}_{0})^2]\nonumber \\ 
& &
-\tilde{\gamma}[2 A^1  (\nabla \cdot A)^{\prime} + (\nabla \cdot A)^2] - \Lambda \} \ .
\label{T33}
\end{eqnarray}

The line element (\ref{LE}) does not depend on time and, consequently, 
it does not describe a cosmological space-time. This is also valid in GR, where the same line element 
leads to various metrics as those of Schwarzschild and Kottler-Schwarzschild-de Sitter [see \cite{kot18}].
The region where these solutions are physically significant 
must be determined in each case; e.g., regions where $g_{00} > 0 $ must be excluded.
In VT, it has been claimed (see above) that the cosmological constant 
is related with the value of $\nabla \cdot A $ in the background universe, but this value is different from 
that of the same divergence in the stationary spherically symmetric case;
by this reason, in spite of its origin, the cosmological constant is treated as 
in GR, and it is denoted $\Lambda$.

Let us now look for the stationary spherically symmetric solutions of the VT 
field equations following 
various steps.

\subsection{First step: proving that $\nabla \cdot A $ is constant}
\label{S_2}

The calculation of $\nabla \cdot A $ may be performed by solving 
the tensor field equation~(\ref{gt}) for $\mu=1$ and $\nu=0$.
In this case, since the components $R^{1}_{\,\,\, 0}$ and $\delta^{1}_{\,\,\, 0}$ vanish, from Eq.~(\ref{TA01}) one 
easily obtains: $2 \tilde{\gamma} g^{00} g^{11}  A_{0} (\nabla \cdot A)^{\prime}=0$; hence, 
for $\tilde{\gamma} \neq 0$,
$g^{00} \neq 0$, $ g^{11} \neq 0$, and
$A_{0} \neq 0$, it follows that $(\nabla \cdot A)^{\prime}$ vanishes and, consequently, 
a trivial integration gives
\begin{equation}
\nabla \cdot A = K_{0} \ ,
\label{divae}
\end{equation}
where $K_{0}$ is an integration constant.

\subsection{Second step: deriving the relation $\alpha(r)=-\beta(r)$}
\label{S_1} 

The trace $T$ is first calculated by using the $T^{\mu \nu} $ components
calculated from Eqs.~(\ref{TA01}) --~(\ref{T33}) and the 
$g_{\mu \nu}$ metric components. The result is 
\begin{equation}
T = - 4 \tilde{\gamma} [  A^1  (\nabla \cdot A)^{\prime} +  (\nabla \cdot A)^2 ] - 4 \Lambda \ ,
\end{equation} 
and, then, taking into account this result, Eq.~(\ref{tg}), and Eq.~(\ref{divae}), one easily get the
relation 
\begin{equation}
R^{0}_{0} = R^{1}_{1}  \ ,
\label{R00R11} 
\end{equation} 
From this equation and the nonvanishing components of the Ricci tensor:
\begin{equation}
R_{00} \ = \ e^{2(\alpha - \beta)} \left [
\alpha^{\prime \prime} + (\alpha^{\prime})^2 - \alpha^{\prime} \beta^{\prime} + 2 r^{-1} \alpha^{\prime}
\right ]  \ ,
\label{R00} 
\end{equation}
\begin{equation}
R_{11} \ = 
- \alpha^{\prime \prime} - (\alpha^{\prime})^2 + \alpha^{\prime} \beta^{\prime} + 2 r^{-1} \beta^{\prime} \ 
\end{equation}
\begin{equation}
R_{22} \ = \ e^{-2\beta} \left [
r (\beta^{\prime}-\alpha^{\prime}) - 1
\right ] + 1 \ ,
\end{equation}
\begin{equation}
R_{33} \ = \ sin^2\theta \ R_{22} \ ,
\label{R33} 
\end{equation}
the following relation is easily obtained:
\begin{equation}
2 r^{-1}(\alpha^{\prime} + \beta^{\prime} ) \ = \ 0 \ .
\end{equation}
The same equation is also obtained in GR. After integration, it leads to $\alpha \ = \ - \beta$
[see, e.g., \cite{ste03}]. Evidently, this relation implies that $g^{00}g^{11}=-1$.

\subsection{Third step: calculation of the $A_0$ component}
\label{S_3}

Function $A_{0}(r)$ may be calculated by solving 
Eq.~(\ref{1.3_vt}) in the stationary spherically symmetric case.
Since $\nabla \cdot A$ has been proved to be constant [see Eq.~(\ref{divae})],
the vector $J^{^{A}}_{\mu} $ vanishes and, consequently, 
taking into account the relation $ 2 \epsilon \neq  \gamma$, 
which must be satisfied (see Sect. \ref{sec-1}),
Eq~(\ref{1.3_vt}) reduces to $\nabla^{\nu} F_{\mu \nu} = 0$.
Moreover, taking into account 
Eqs.~(\ref{VF})--(\ref{nvf}), the 
covariant derivative $\nabla^{\nu} F_{\mu \nu}$ may be 
easily calculated to get
\begin{equation}
A^{\prime \prime}_0 + 2 r^{-1} A^{\prime}_0 \ = \ 0 \ .
\label{EqA0}
\end{equation}
\noindent
In terms of the new variable $y = A^{\prime}_0$, the last equation reduces to 
$y^{\prime}+2 r^{-1} y= 0$. The solution of this equation is 
$y=A^{\prime}_0=-R_0/r^2$ and, then, a new integration gives
\begin{equation}
A_0(r) \ = \ R_0 r^{-1} + R_1 \ ,
\label{solA0} 
\end{equation}   
\noindent
where $R_0$ and $R_1$ are integration constants.

\subsection{Fourth step: computing metric components}
\label{S4}

For $\mu=1 $ and $\nu = 1 $, the tensor field equation (\ref{tg}) may be easily
written in the form 
\begin{eqnarray}
(\tilde{\gamma} - 2 \tilde{\varepsilon})  (A^{\prime}_{0})^2 - \tilde{\gamma} (\nabla \cdot A)^2 - \Lambda - T/2 = \nonumber \\ 
e^{2\alpha} \ (- \alpha^{\prime \prime} - 2 (\alpha^{\prime})^2  - 2 r^{-1} \alpha^{\prime} ) \ ,
\label{tg11}
\end{eqnarray}
\noindent
where we have taken into account the relation $g^{00}g^{11} = -1$ (see Sect.~\ref{S_2}),
the nonvanishing components of $R_{\mu\nu}$ and $T^{\mu\nu}$ listed in previous Sections, 
and Eq.~(\ref{divae}).
In the same way, for $\mu=2 $ and $\nu = 2 $, one finds
\begin{eqnarray}
(2 \tilde{\varepsilon} - \tilde{\gamma} )  (A^{\prime}_{0})^2 - \tilde{\gamma} (\nabla \cdot A)^2 - \Lambda - T/2 =\nonumber \\ 
\frac{1}{r^2} [e^{2 \alpha} (-2 \alpha^{\prime} r - 1) + 1] \ .
\label{tg22}
\end{eqnarray}

Subtracting equations (\ref{tg11}) and (\ref{tg22}) 
and multiplying by the factor $e^{-2 \alpha}$, the following second order differential equation
is obtained:
\begin{equation}
\alpha^{\prime \prime} + 2 (\alpha^{\prime})^2  + [ r^{-2} + 2 (\tilde{\gamma} - 2 \tilde{\varepsilon})  (A^{\prime}_{0})^2 ]
e^{-2\alpha} - r^{-2} \ = \ 0    \ .
\label{EDOa}
\end{equation}
This equation can be solved by using the new variable $w \ = \ e^{2\alpha}$. In terms of $w$,
Eq.~(\ref{EDOa}) reads as follows:
\begin{equation}
w^{\prime \prime} - 2 r^{-2} w \ = \ g(r)  \ ,
\label{EDOb}
\end{equation}                       
\noindent
where $g(r) = -2 [ r^{-2} + 2 (\tilde{\gamma} -2 \tilde{\varepsilon}) A^{\prime}_{0})^2 ] $ and
$A^{\prime}_0 = -R_{0}/r^{2}$ (see above in this Section).
The general solution of Eq.~(\ref{EDOb}) is $w=w_h + w_p$, where $w_h$
is the general solution of the
corresponding homogeneous equation, and $w_p$ is a particular solution
of the complete inhomogeneous equation. The general solution $w_h$ is:
\begin{equation}
w_h \ = \  C_1 w_1(r) + C_2 w_2(r) \ = \ C_1 r^2 + C_2 r^{-1} \ ,
\label{WH}
\end{equation}
\noindent
$C_1$ and $C_2$ being integration constants.

In order to obtain a particular solution, $w_p$, we may apply the method of 
parameter variations; according to this method, we must look for a solution 
of the following form:
\begin{eqnarray}
w_p \ = \  u_1(r) w_1(r) + u_2(r) w_2(r) \ = \nonumber \\ 
\ u_1(r) r^2 + u_2(r) r^{-1},
\end{eqnarray}
\noindent
where
\begin{eqnarray} 
u_1(r) & = & - \int \frac{w_2(r) g(r)}{W_{(w_1,w_2)}(r)} \ dr \nonumber \\
& &  \nonumber \\
u_2(r) & = & \int \frac{w_1(r) g(r)}{W_{(w_1,w_2)}(r)}  \ dr \ ,
\end{eqnarray}
\noindent
and $W_{(w_1,w_2)}(r)$ is the Wronskian:
\begin{eqnarray}
W_{(w_1,w_2)}(r) \ = \
\left |
\begin{array}{c c}
w_1(r) & w_2(r)  \\
w^{\prime}_1(r) & w^{\prime}_2(r) 
\end{array}
\right | \ = \nonumber \\ 
\ w_1(r) w^{\prime}_2(r) - w^{\prime}_1(r) w_2(r) \ = \ - 3.
\end{eqnarray}
So, the particular solution $w_p$ takes on the form:
\begin{eqnarray}
w_p(r) \ = \ 1 - \frac{4}{3} \ (\tilde{\gamma} -2 \tilde{\varepsilon}) [
r^2 \int r^{-1} (A^{\prime}_0(r))^2 dr -\nonumber \\ 
r^{-1} \int r^2 (A^{\prime}_0(r))^2 dr ] \ .
\label{wp}
\end{eqnarray}
Let us now use the explicit form of $A^{\prime}_0$ (see above) to easily find
\begin{equation}
w_p(r) \ = \ 1 + (2 \tilde{\varepsilon} - \tilde{\gamma}) \ R_0^2  \ r^{-2} \ .
\label{WPR}
\end{equation}

Finally, Eqs.~(\ref{WH}) and~(\ref{WPR}) allow us to write the general form 
of function $w=e^{2\alpha} $, which directly leads to the metric components
\begin{equation}
g_{00} \ = \ - \big[ 1 + C_1 r^2 + C_2 r^{-1}  + (2 \tilde{\varepsilon} - \tilde{\gamma}) R_{0}^{2} \ r^{-2}\big] \ ,
\label{g00}
\end{equation}     
and 
\begin{equation}
g_{11} \ = -g_{00}^{-1} \ .
\label{g11}
\end{equation}

\subsection{Fifth step: calculation of the $A_1$ component}
\label{S_5}

The last step is the integration of Eq.~(\ref{divae}) --derived in Sect.~\ref{S_2}-- 
to get the function $A_{1}(r)$. This equation may be easily rewritten as follows
\begin{equation}
\nabla \cdot A \ = e^{2 \alpha} \left [ 2 (\alpha^{\prime} + r^{-1} ) A_1 + A^{\prime}_1
\right ] \ = \ K_0 \ .
\label{div}
\end{equation}
This is a linear first order differential equation of the form $h(r) A^{\prime}_1 = f_1(r) A_1 + K_0$, 
with $h(r)=e^{2 \alpha}$ and $f_1(r)=- 2 e^{2 \alpha} (\alpha^{\prime}+r^{-1} )$. The solution 
of this equation is:
\begin{equation}
A_1(r) = e^{F(r)} \left ( K_1 + \int e^{-F(r)} \frac{K_0}{h(r)} \ dr
\right ) \ , 
\end{equation}
where $K_1$ is another integration constant, and   
$F(r) \ = \ \int [f_1(r)/h(r)] \ dr $.  
After performing these integrals, one obtains:
\begin{equation}
A_1(r) = \frac{ K_0 r/ 3 +  K_1 r^{-2} } 
{1 + C_1 r^2 + C_2/r
+ (2 \tilde{\varepsilon} - \tilde{\gamma}) R_0^2 \  r^{-2}} \ .
\label{solA1}
\end{equation} 

Eqs.~(\ref{solA0}) and~(\ref{solA1}) give the vector field  $A^{\mu} $, and 
Eqs.~(\ref{g00}) and~(\ref{g11}) define the metric of the VT theory in the stationary spherically symmetric 
case. The resulting metric is a generalization of the 
Kottler-Schwarzschild-de Sitter 
one, which is obtained for $C_1= - \frac{\Lambda}{3}$, $C_2=- 2 m =- R_S$ 
($R_S$ being the Schwarzschild radius), and 
$(2 \tilde{\varepsilon} - \tilde{\gamma}) =0$. In the VT theory we have found a  new term 
$ (2 \tilde{\varepsilon} - \tilde{\gamma})  R_0^2/r^{2}$, which is positive
due to the fact that the relation  $2 \epsilon - \gamma > 0$ must be satisfied
[see Sect.~\ref{sec-1}]. The metric obtained in the framework of the VT theory is similar to 
the Reissner-Nordstr\"{o}m-de 
Sitter metric (\cite{kay79}), which corresponds to a stationary spherically symmetric charged system in GR.
The form of this known metric is
\begin{equation}
g_{00}=-g_{11}^{-1}= -\Big[1-\frac{2m}{r}-\frac {\Lambda}{3}r^{2} + \frac {Q^{2}}{r^{2}}\Big] \ ,
\end{equation}
it involves a positive term proportional to $1/r^{2} $ which depends on the 
electrical charge $Q$; evidently, 
in Eqs.~(\ref{g00}) and~(\ref{g11}), there is also a term of this kind, in which,
the role of $Q^{2}$ is played by the constant $(2 \tilde{\varepsilon} - \tilde{\gamma})  R_0^2$.

\section{Horizons in the VT theory}   
\label{sec-4}

In the stationary spherically symmetric case, outside the matter distribution, and in the
absence of electrical charge, the solution of the VT 
field equations involves the integration constants $R_{0} $, $R_{1} $, $K_{0} $, $K_{1} $,
$C_{1} $, and $C_{2} $. In this situation, 
the physical system under consideration is fully described by the quantities $m$ and $\Lambda $, whose 
dimensions --in geometrized units-- are $L^{1} $ and $L^{-2} $, 
respectively. Let us now perform a dimensional analysis to predict the dependence
of the parameter $R_{0} $ involved in the metric components in terms of 
$m$ and $\Lambda$.

The constants $C_{1} $, and $C_{2} $ also appear in GR. Since the dimensions of 
$g_{\alpha \beta}$ are $L^{0} $, the term $C_{2}/r$ involved in $g_{00} $ is 
dimensionless and, consequently, the dimension of $C^{2} $ must be $L^{1} $; hence, $C_{2} $ must be 
a dimensionless number, $\tilde{C}_{2} $, multiplied by $m$; in this case, we have a well known 
criterion to 
conclude that $\tilde{C}_{2} = -2$, a number leading to the well known term
$-2m/r $. In the same way,
the dimensionless character of $C_{1}r^{2}$ leads to the conclusion that 
the dimension of $C_{1} $ is $L^{-2} $; hence, this term must be the product of a 
dimensionless constant $\tilde{C}_{1} $ by the factor $\Lambda $. In this case, 
there are also arguments to conclude that $\tilde{C}_{1} = -1/3$, a number 
which leads to the well known term $-\Lambda r^{2}/3 $ involved in the 
Kottler-Schwarzschild-de Sitter metric.

A similar analysis may be performed for the constants $R_{0} $ and $R_{1}$; in fact, 
according to Eq.~(\ref{solA0}), the dimensionless component 
$A_0 $ is the sum of two terms of the form  $R_0 r^{-1}$ and  $R_1 $;
hence, the dimensions of $R_{0} $ and $R_{1} $ are $L^{1} $ and $L^{0} $,
respectively and, consequently, we conclude that the constant $R_{0} $ must be the product 
of a dimensionless constant 
$\tilde{R}_{0} $ by $m$; in this case, we have not any criterion to fix the 
dimensionless constant $\tilde{R}_{0} $, which keeps arbitrary by the moment.
This analysis does not give any information about 
the dimensionless constant $R_{1} $, but this information is not necessary 
to look for the horizons, which follows from the fact that 
--according to Eqs.~(\ref{g00}) and (\ref{g11})-- 
the metric components do not depend on $R_{1} $.

The dimensional analysis in not extended to 
the component $A_1 $, since the metric is also 
independent of the constants $K_{0} $ and $K_{1} $ involved in 
Eq.~(\ref{solA1}).

After the above dimensional considerations we can write:
\begin{equation}
A_0(r) \ = \ \tilde{R}_0 m/r \ ,
\label{solA02}
\end{equation} 
\begin{equation}
g_{00} \ = \ -g_{11}^{-1}= - \Big[1  - \frac {\Lambda}{3} r^2 - \frac {2m} {r} 
+ (2 \tilde{\varepsilon} - \tilde{\gamma}) 
\tilde{R}_0^2 \frac {m^{2}} {r^{2}}\Big] \ ,
\label{g002}
\end{equation}     
where $\tilde{R}_0 $ plays the role of a dimensionless arbitrary constants,
which should be fixed by studying appropriate nonlinear problems in the 
framework of the VT theory, as, e.g., the geodesic motion of proof particles 
close to possible horizons.

In terms of the function $f(r) = - g_{00}(r)$,
the horizons are the hypersurfaces $r=r_{h} $ defined by the condition $f(r_{h}) =  0$.
In the regions where the inequality $f(r) > 0$ is satisfied, our description 
of the stationary spherically symmetric space-time is physically consistent.
Condition $f(r) < 0$ is not compatible with the assumed metric signature.

In the standard $\Lambda$CDM cosmological model of GR, most current observations are explained 
for values of the vacuum energy density parameter $\Omega_{\Lambda} $ 
close to $ 0.73 $, which corresponds to $\Lambda \simeq 10^{-46} \ Km^{-2} $. The same value 
also explains current observations in the framework of the VT theory [see \citet{dal14}]; hence,
the above 
value of the cosmological constant is hereafter fixed.

The mass $m$ is varied between $10 \ M_{\odot} $ and $10^{9} \ M_{\odot} $; so, the masses of
different types of black holes are considered. From stellar black holes due to supernova
explosions, to
supermassive ones located in the galactic central regions.

Once a mass $m$ has been fixed, function $f(r) $ only involves 
the unknown positive parameter $\Gamma = (2 \varepsilon - \gamma)
\tilde{R}_0^2 $. For $\Gamma = 0$, the metric reduces to the 
Kottler-Schwarzschild-de Sitter one and, in such a case, there 
are two horizons, the first (second) one is the black hole (cosmological) 
horizon, whose radius is hereafter denoted $r_{BH} $  ($r_{C} $).
In the region limited by these two horizons, namely, for  $r_{BH} < r < r_{C} $,
function $f(r) $ is positive and the Kottler-Schwarzschild-de Sitter 
metric is physically admissible.

\citet{kay79} studied the horizons in the 
Reissner-Nordstr\"{o}m-de Sitter space-time. 
If the outcomes obtained in that paper are rewritten in
our case, by replacing $Q^{2} $ by $(2 \tilde{\varepsilon} - \tilde{\gamma}) m^{2} \tilde{R}_0^2$,
it is straightforward to conclude that
from $\Gamma=0 $ to a certain $\Gamma$ value, $\Gamma_{max} $, which 
is greater than $1/8\pi $ but very close to it, 
there are both a black hole horizon and a cosmological one; 
however, for $\Gamma > \Gamma_{max} $, 
there is an unique horizon which is cosmological.
Our calculations have verified all this.

For appropriate $m$ values, the algebraic equation $f(r)=0$ has been 
numerically solved for $\Gamma =0$ and for many positive $\Gamma $ values. 
For $\Gamma > 1/8\pi $ and 
whatever $m$ may be, we have found only a root at $r_{C} \simeq 1.73 \times 10^{23} \ Km $
(there is no black hole horizon). However, for any $\Gamma < 1/8\pi $, apart from the 
above $r_{C} $ radius for the cosmological horizon, we have also obtained 
a black hole horizon with a $r_{BH} $ radius depending on both $m$ and 
$\Gamma $. 

\begin{figure*}[tb]
\begin{center}
\resizebox{1.\textwidth}{!}{%
\includegraphics{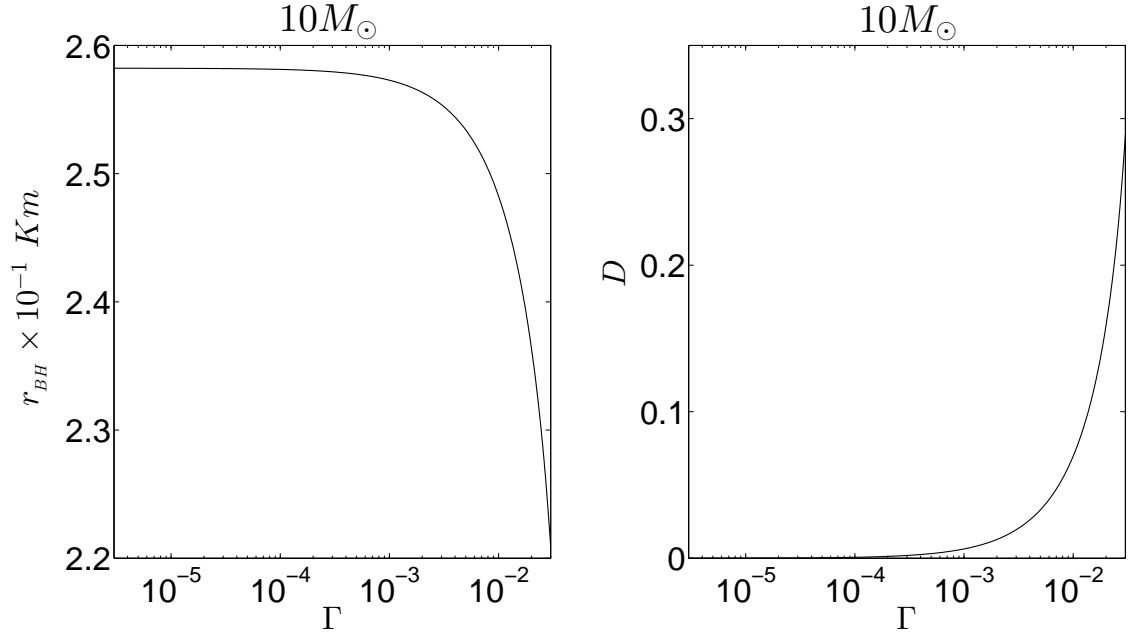}
}
\caption{Left panel shows the radius 
$r_{BH}$ of the black hole horizon as a function of $\Gamma $.
Right panel represents the relative deviations, $D$, between 
$r_{BH}(\Gamma=0)$ and $r_{BH}(\Gamma)$, in the $\Gamma $ interval
where there is a black hole horizon. In both panels $m = 10 \ M_{\odot} $}
\label{figu1} 
\end{center}      
\end{figure*}

Figure~\ref{figu1} corresponds to a mass $m = 10 \ M_{\odot} $ (stellar black hole).
The left panel shows $r_{BH} $ as a function of $\Gamma $ inside the 
interval [0, $1/8\pi$]. The radius of the black hole horizon decreases
as $\Gamma $ separates from the zero value corresponding to the 
Kottler-Schwarzschild-de Sitter solution of the GR field equations.
In the right panel, the relative deviation 
\begin{equation}
D= \frac {2[r_{BH}(\Gamma=0)-r_{BH}(\Gamma)]}
{r_{BH}(\Gamma=0)+r_{BH}(\Gamma)}
\label{drr}
\end{equation}
is represented, as a function of $\Gamma $, in 
the same interval as in the left panel. We see that these deviations 
reach values close to 30\%, which are not very large deviations, but 
moderate significant ones.

\begin{figure*}[tb]
\begin{center}
\resizebox{1.\textwidth}{!}{%
\includegraphics{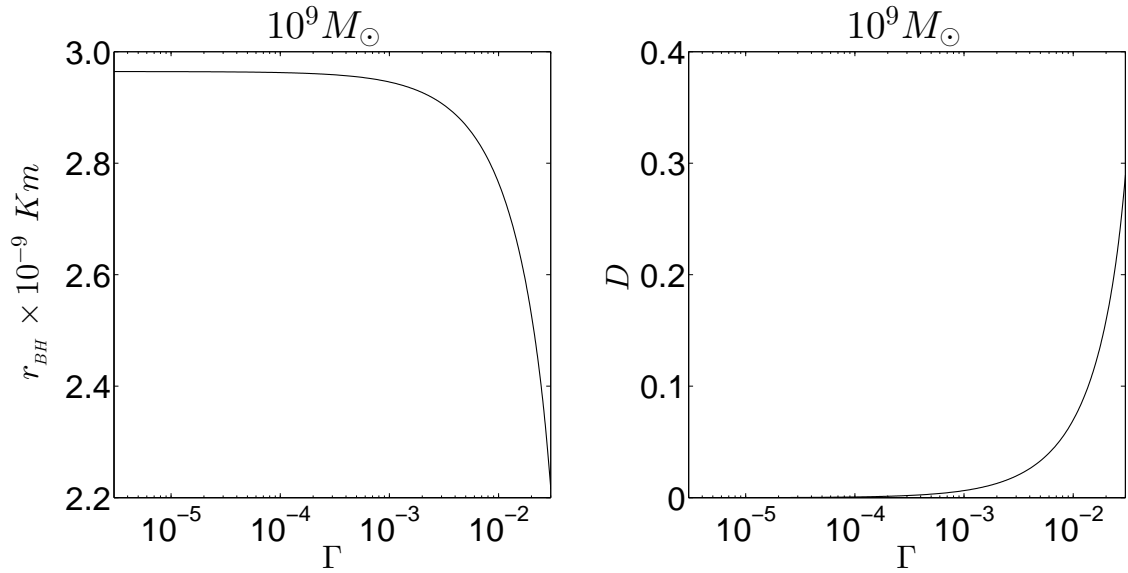}
}
\caption{Same as in Fig.~\ref{figu1} for $m=10^{9} \ M_{\odot}$}
\label{figu2} 
\end{center}      
\end{figure*}

In Figure~(\ref{figu2}), the mass is $m = 10^{9} \ M_{\odot} $ (galactic
supermassive black hole) and, consequently, 
the radius $r_{BH}(\Gamma=0) \simeq 2m $ is greater than in the 
top panels by a factor of $10^{8} $; nevertheless, this proportionality factor is 
the same for any $\Gamma$ and, consequently, the form of the curves represented in the 
left panels of Figs.~(\ref{figu1}) and~(\ref{figu2}) are identical.
Moreover, the relative deviations $D$ defined in Eq.~(\ref{drr}) 
reach the same values in the 
right panels of the two Figures, which means that these deviations 
do not depend on $m$.

\section{Conclusions.}   
\label{sec-5}

This paper has been devoted to the 
development of the VT theory of gravitation proposed by \citet{dal09}.
Previous applications of this theory to both the solar system and 
cosmology have given excellent results \citep{dal14}. 
Here, we have solved the field equations 
of the VT theory, in the absence of matter and electrical charge,
by assuming a stationary spherically symmetric space-time.
It has been proved that the resulting solution has the same form 
as the Reissner-Nordstr\"{o}m-de 
Sitter solution of GR, but the role of the electrical charge is played
by a quantity proportional to the source mass $m$.
After reaching this conclusion, we have focused our attention 
on the horizons associated to stellar and massive black holes. 
 
\citet{noj14} have proved that, 
in the absence of electrical charge,
there are 
$f(R) $ theories of gravitation leading to 
Reissner-Nordstr\"{o}m-de Sitter space-times, but   
the authors recognize that -in these theories-- the meaning of the quantity playing the role of 
the electrical charge is not clear.   

Since the cosmological constant, $\Lambda$, is fixed by comparisons between 
predictions of the VT theory and current observations, 
the cosmological horizon is practically constant. Its radius is almost 
independent of the mass, $m$, for any realistic black hole. 
There is always a cosmological horizon whatever the value of the parameter $\Gamma $ 
defined in Sect.~\ref{sec-4} may be.

In the VT theory, we have proved that, for a given mass $m$,  
the radius of the black hole horizon is smaller than    
the radius of the Kottler-Schwarzschild-de Sitter
black hole having the same mass.
The relative deviations between these two radius 
are small but significant, reaching values close to 30 \%.
This effect is important since it is similar to the effect 
due to the black hole rotation in GR, which leads to a 
horizon radius smaller than that corresponding to $J=0$. 
For $\Gamma > 1/8\pi $ there is no any black hole 
horizon in the VT theory under consideration.

Various methods have been designed to estimate the mass $m$ and angular 
momentum $J$ of a black hole from observations. 
If, in future, the mentioned methods become accurate enough, 
and the estimated $m$ and $J$ quantities obey the relation 
predicted by means of the Kerr solution of Einstein equations,
the contribution of the vector field to the horizon radius
will have to be considered negligible ($\Gamma \simeq 0$); however, 
if the Kerr relation is not satisfied by the observed 
values of $m$ and $J$, an appropriate  $\Gamma $ 
value could solve the problem.

Let us finally mention two interesting extensions of this paper:
first of all, the motion of test particles in the neighborhood of 
the above VT black hole deserves attention; so, accretion disks 
and other phenomena might be studied. Afterward,
the stationary axially symmetry line element, plus an
appropriate vector field $A^{\mu}$, should be considered
to study rotating black holes in the framework of the VT 
theory; in this way, a relation between $m$ and $J$ could be 
found, which might be satisfied by accurate future 
observed values of these quantities.

\acknowledgments{This research has been supported by the Spanish
Ministry of {\em Econom\'{\i}a y Competitividad},
MICINN-FEDER project FIS2012-33582}

\end{document}